# Generative AI as a Tool for Enhancing Reflective Learning in Students

Bo Yuan
The University of Queensland
Brisbane, Australia
boyuan@ieee.org

Jiazi Hu
AI Consultancy
Kunming, P.R. China
jiazi841231@163.com

***Abstract*—Reflection is widely recognized as a cornerstone of student development, fostering critical thinking, self-regulation, and deep conceptual understanding. Traditionally, reflective skills have been cultivated through structured feedback, mentorship, and guided self-assessment. However, these approaches often face challenges such as limited scalability, difficulties in delivering individualized feedback, and a shortage of instructors proficient in facilitating meaningful reflection. This study pioneers the use of generative AI, specifically large language models (LLMs), as an innovative solution to these limitations. By leveraging the capacity of LLMs to deliver personalized, context-sensitive feedback at scale, this research investigates their potential to serve as effective facilitators of reflective exercises, sustaining deep engagement and promoting critical thinking. Through in-depth analyses of prompt engineering strategies and simulated multi-turn dialogues grounded in a project-based learning (PBL) context, the study demonstrates that, with pedagogically aligned prompts, LLMs can serve as accessible and adaptive tools for scalable reflective guidance. Furthermore, LLM-assisted evaluation is employed to objectively assess the performance of both tutors and students across multiple dimensions of reflective learning. The findings contribute to the evolving understanding of AI's role in reflective pedagogy and point to new opportunities for advancing AI-driven intelligent tutoring systems.**

***Keywords—Generative AI, Large Language Models, Reflective Learning, AI in Education, Intelligent Tutoring Systems***

## I. Introduction

Reflective learning, a process through which learners critically assess their experiences to gain insights and improve future performance, holds a central role in educational theory and practice [1]. Defined as a deliberate act of self-assessment and analysis, reflection promotes a deeper understanding of both academic content and personal growth by encouraging students to question assumptions, evaluate outcomes, and engage in continuous improvement. A key element of this process is self-distancing, the ability to step back from one's experiences and analyze them from a more objective perspective. This skill allows learners to identify patterns, recognize strengths and weaknesses, and refine their approaches to learning [2]. Reflective learning is often facilitated through methods such as journaling, self-assessment exercises, and guided discussions, which provide structured opportunities for students to articulate and develop their thoughts and insights. By fostering critical thinking, self-awareness, and the ability to transfer knowledge to new contexts, reflective learning serves as an indispensable component of transformative education.

Despite its well-recognized value, implementing reflective learning in traditional educational environments poses several challenges. A primary obstacle is the demand for sustained, individualized feedback to effectively guide students through the reflective process. In typical classroom environments, where instructors often manage large groups of students, providing personalized feedback on each learner's reflections can be both time-intensive and logistically demanding. This constraint limits the opportunities for students to receive timely and specific insights, reducing the depth and frequency of reflective engagement. Furthermore, facilitating reflective learning requires specific expertise in fostering self-examination and critical analysis, which may not be adequately and consistently developed among educators. These challenges underscore the urgent need for scalable and accessible tools to support and enhance reflective learning, enabling its broader adoption and impact in diverse educational contexts.

The advent of generative artificial intelligence (AI), particularly large language models (LLMs) [3], has opened new avenues for delivering effective, natural language feedback in educational contexts [4]. Capable of generating nuanced, context-sensitive responses, LLMs can emulate the role of an instructor, offering guidance and feedback that closely resembles human interaction. This ability to deliver timely and personalized feedback positions generative AI as a transformative tool in modern education, offering innovative solutions to the challenges inherent in reflective learning.

This study leverages LLMs to address limitations in traditional reflective practices by providing accessible, personalized guidance that fosters deep student engagement. Through structured prompt engineering and tailored model responses, these AI systems can effectively guide students through the reflective process, encouraging them to explore alternative perspectives, reassess assumptions, and uncover new insights in a supportive environment. Additionally, LLMs possess the ability to analyze written reflections, discussions, and other learning artifacts, providing constructive feedback on critical thinking, self-awareness, and the depth of reflection. By identifying meaningful patterns and trends, they can complement traditional assessment methods, bridging the gap between subjective interpretation and data-driven insights. Notably, the integration of LLMs marks a major advancement in intelligent tutoring systems (ITSs) [5, 6], extending their role beyond traditional content delivery and assessment to include active support for reflective learning practices.

This paper is structured as follows. Section II provides a concise review of related work, highlighting foundational concepts in reflective learning and intelligent tutoring systems. Section III discusses the design and optimization of prompts for interactive reflective learning, underscoring the importance of well-crafted prompts in eliciting meaningful and thoughtful student responses. Section IV evaluates the performance of LLMs, with GPT-4o as a case study, in supporting reflective learning, focusing on their capacity to guide learners toward deep self-assessment and critical reflection. Section V concludes the paper by summarizing the key findings and outlining future directions for advancing the role of AI in reflective learning.

## II. Related Work

### A. Reflective Learning

The Gibbs Reflective Cycle is a structured framework for reflective learning [7], which provides a systematic approach to analyzing learning experiences (Fig. 1). It consists of six stages: the *Description* stage, where the details of the event are recalled; the *Feelings* stage, which explores the emotions experienced by the individual and others during the event; the *Evaluation* stage, where both the positive and negative aspects of the experience are assessed; the *Analysis* stage, where the causes and patterns behind the outcomes are explored in depth; the *Conclusion* stage, which summarizes lessons learned and potential improvements; and the *Action Plan* stage, which creates specific strategies for applying these insights in future situations.

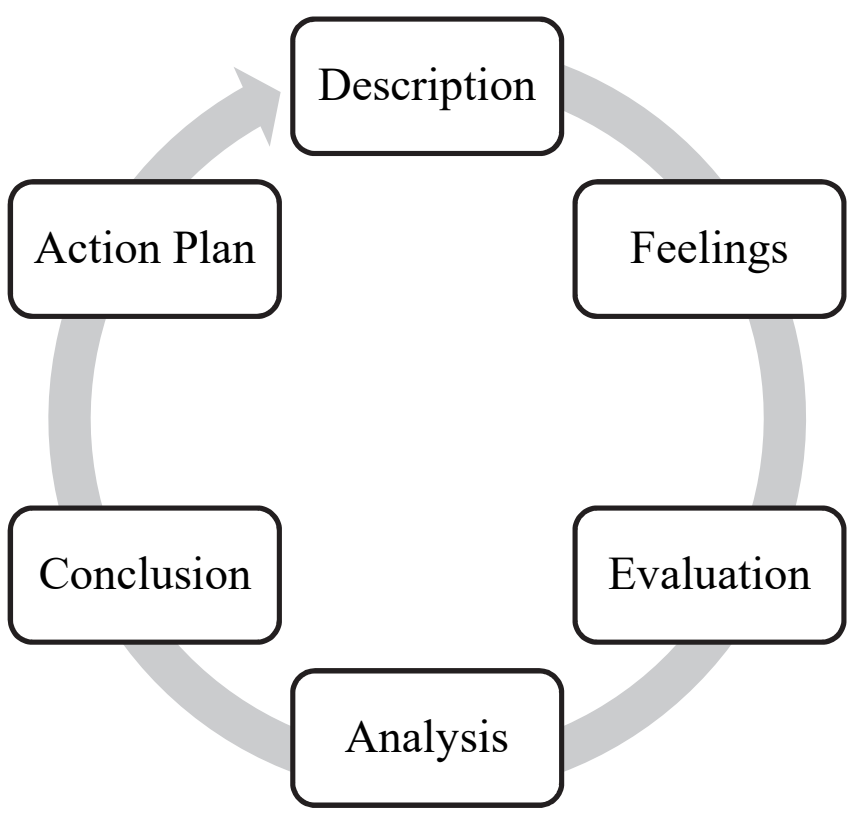

Fig. 1. The Gibbs Reflective Cycle. This six-stage framework supports reflective learning by guiding individuals through description, feelings, evaluation, analysis, conclusion, and action planning. It provides a systematic approach to evaluating experiences, deriving meaningful insights, and formulating strategies for improvement.

Reflective learning encourages students to revisit and analyze their learning experiences, emotional responses, decision-making processes, and outcomes. Traditional approaches to reflective learning include learning journals, reflection diaries, group discussions, peer feedback, and teacher guidance. However, these approaches face several challenges in practice. For example, many students are not accustomed to engaging in deep reflection and may focus only on the surface-level aspects of their learning, lacking thorough analysis of their experiences. Meanwhile, some students may be reluctant to reflect openly due to cultural or psychological factors, such as low self-confidence or fear of exposing their weaknesses.

To maximize the effectiveness of reflective learning, LLMs can guide students through structured reflection processes. For instance, LLMs are able to provide continuous and tailored scaffolding that encourages students to analyze their learning experiences in greater depth, thereby fostering reflection that goes beyond surface-level observations. They can also create a safe, non-judgmental environment in which students feel comfortable articulating their thoughts and emotions without fear of embarrassment or criticism. In addition, LLMs can deliver constructive and affirming feedback that highlights the value of reflections and suggests directions for further exploration. Table I summarizes the key dimensions through which LLMs may be integrated into reflective learning.

TABLE I. Integration of LLMs in Reflective Learning

| Dimension | Benefits for Reflective Learning |
|---|---|
| Prompt Generation | LLMs can generate dynamic, context-specific prompts informed by a student's prior responses. |
| Real-Time Feedback | LLMs can provide instant feedback on reflections, highlighting areas for deeper analysis and suggesting improvements. |
| Personalized Learning | LLMs can track individual student progress and deliver tailored guidance aligned with their unique needs and learning styles. |
| Collaborative Reflection | LLMs can facilitate peer feedback by summarizing student reflections, enabling more focused and meaningful feedback. |

### B. Intelligent Tutoring Systems

An ITS is a computer-based instructional platform designed to deliver personalized guidance, adaptive feedback, and tailored experiences, thereby supporting self-directed and autonomous learning [8]. As illustrated in Fig. 2, a typical ITS comprises four interrelated core components. The *Domain Module* stores the knowledge representation of the target subject area, often structured as rule-based expert systems or knowledge graphs. The *Student Module* continuously monitors the learner's progress, performance, and misconceptions to achieve fine-grained learner modeling. The *Tutoring Module* orchestrates instructional strategies by selecting appropriate tasks, generating adaptive hints, explanations, or corrective feedback, and sequencing learning activities to optimize engagement and mastery. The *User Interface* provides the medium for learner–system interaction, ranging from text-based input to multimodal interfaces that support speech or interactive visualizations [9].

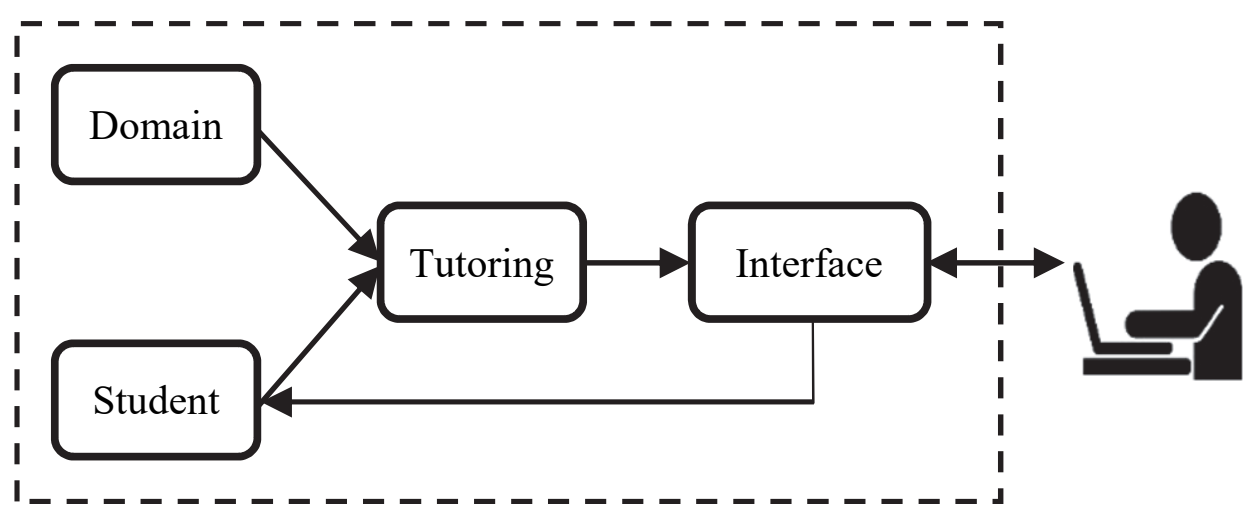

Fig. 2. The functional diagram of an ITS, illustrating four key components and their relationships. The student module continuously updates based on learner input; the tutoring module consults both the domain and student modules to generate personalized feedback; the interface functions by delivering adaptive guidance to the learner and collecting learner responses for system updates.

This emphasis on interaction and feedback aligns closely with the core affordances of LLMs, offering opportunities for deeper engagement and more responsive feedback in intelligent tutoring contexts [10]. Importantly, LLMs possess the capacity to unify ITS modules, reducing the need for explicitly engineered, standalone components. Their integration into ITSs paves the way for a seamless, conversational, and adaptive tutoring experience, enabling systems to fluidly manage content delivery, learner modeling, and feedback generation within a cohesive framework. Table II summarizes the potential impact of LLMs on the core modules of traditional ITSs.

TABLE II. POTENTIAL IMPACT OF LLMS ON CORE MODULES OF ITSS

| Module | Impact with LLM Integration |
|---|---|
| Domain | LLMs provide broad and dynamic knowledge representations, reducing reliance on predefined content and enhancing flexibility in knowledge coverage and acquisition. |
| Student | LLMs analyze learners' natural language input to infer comprehension levels and misconceptions in real time, thereby supporting more adaptive and personalized learning. |
| Tutoring | LLMs deliver pedagogically aligned guidance by generating diagnostic feedback and clarifying misconceptions, thereby enhancing instructional effectiveness. |
| Interface | LLMs enable intuitive, conversational interfaces that support fluid, natural language interactions for enhanced accessibility. |

The research on LLM-enhanced ITSs remains at a relatively early stage [11]. In the specific context of reflective learning, existing studies have primarily investigated applications such as supporting reflective writing [12] or facilitating brief after-class reflections [13]. However, there is still a notable gap in rigorous empirical analyses regarding the efficacy of LLMs in multi-turn dialogues that actively scaffold and sustain students' reflective processes. To address this gap, this work undertakes a systematic examination of the role of LLMs in dialogue-driven reflective learning. In this setting, LLMs are designed to simulate the behavior of experienced tutors, engaging students in structured and interactive conversations to inspire deeper levels of self-reflection and enhance their understanding.

## III. EXPERIMENT SETUP

As outlined in Table II, LLMs can function as effective reflective learning assistants by adapting to students' individual proficiency levels and preferred learning styles. In addition, by virtue of their extensive and interdisciplinary knowledge base, LLMs are well positioned to support cross-domain learning and critical thinking, encouraging students to draw meaningful connections across diverse subject topics. Moreover, their consistent, patient, and unbiased nature can mitigate the risks of emotional fluctuation or subjective bias, fostering a supportive and inclusive learning environment.

This study adopts project-based learning (PBL) [14] as a case study, which contrasts with traditional knowledge-oriented learning by emphasizing hands-on projects rather than the mastery of isolated knowledge points. In PBL, students actively engage in tasks that demand critical thinking, collaboration, and creativity. Reflective learning is therefore integral to PBL, as it allows students to consolidate and refine their understanding through self-assessment and iterative cycles of improvement.

### A. Prompt Design

The LLM prompt in this study was designed in accordance with established prompt engineering guidelines [15] and aligns closely with the core principles of reflective learning. The LLM was configured in a *self-play* mode, simultaneously simulating both the tutor's guidance and the student's responses. This setup enabled the generation of multiple learning sessions without the direct involvement of human participants. By doing so, this study effectively minimized the biases and variability typically associated with human factors, thereby supporting more objective, reliable, and consistent evaluation outcomes.

- *As a tutor with rich experience in reflective learning, your task is to guide a student through reflective learning via multi-turn dialogue. The student recently finished a group project on designing smart devices for campus environmental protection. During the dialogue, you also need to play the role of the student, providing realistic responses that reflect a student's perspective.*
- *First, ask the student to reflect on this experience, identifying one challenge he and his team overcame and one they couldn't overcome.*
- *Then, ask: "Based on your reflections on these challenges, what new insights have you gained about project-based learning?"*
- *Use open-ended questions, encouraging the student to clarify key ideas in detail. Acknowledge his responses with active listening cues, such as "That's an interesting point." or "I see how that was challenging."*
- *Actively follow up with questions that deepen reflection. For example, if the student says he gained a new understanding of an issue, ask him to compare it with his previous perspective and explain how he arrived at this new insight.*
- *Request specific examples. If the student mentions a shift in views, prompt him for examples from his experience that illustrate this change.*
- *Highlight when his reflections show particular depth or progress, noting if there's a growth in thinking or perspective. Encourage the student to consider how these insights might apply to future projects.*

### B. Analysis of Prompt

By integrating open-ended questions, structured prompts, and active listening, this prompt fosters a supportive environment that encourages students to reflect on their experiences, derive meaningful insights, and transfer them to future learning contexts. A detailed breakdown and analysis of the prompt are presented below:

- *Role Setting and Task Definition*: The prompt begins by defining the instructor's role as an experienced tutor in reflective learning. It instructs the tutor to guide the student's reflection while also playing the student's role.
- *Context Setting and Background Information*: It provides concise background information about the student. This helps the tutor understand the project focus and tailor questions to the student's specific experiences, making the reflective process more relevant and grounded.

- *Step-by-Step Reflective Prompts*: The reflection process is broken into structured steps. The student is first asked to identify one challenge that was successfully overcome and one that remained unsolved. This dual focus on successes and struggles encourages balanced reflection, which is critical to meaningful reflection.
- *Encouragement of Insight Development*: The tutor is guided to ask questions such as "What new insights have you gained about project-based learning?" Such prompts move students beyond mere recall, supporting the development of higher-order reflection that links their experiences to broader learning principles.
- *Use of Open-Ended Questions and Active Listening Cues*: Open-ended questions, combined with active listening responses (e.g., "I see how that was challenging"), validate student contributions and foster psychological safety. This creates conditions for deeper sharing, more nuanced reflection, and stronger tutor-student rapport.
- *Encouraging Deeper Reflection through Follow-Up*: The tutor is advised to ask students to compare new insights with their prior perspectives. This follow-up promotes critical thinking by encouraging students to articulate how their understanding has evolved and to become more aware of their learning processes.
- *Requesting Specific Examples*: When students mention a shift in perspective, the tutor prompts them to provide concrete examples. This strategy grounds abstract reflections in authentic experiences, enhancing both the credibility and depth of reflection.
- *Highlighting Growth and Future Application*: Finally, the prompt encourages the tutor to acknowledge progress in student reflections and to guide them in considering how these insights can inform future projects, making reflection both descriptive and transformative.

## IV. Experiments

### A. The Overall Performance

In the experiments, GPT-4o was employed as a reflective learning assistant, functioning on the basis of the designed prompt. Five independent dialogue sessions were conducted and recorded for subsequent analysis. Table III presents the relationships among key concepts relevant to reflective learning, such as project-based learning with team collaboration, technical skills for idea implementation, and reflection for future project improvement.

In terms of text statistics, each session typically comprised around 5 dialogue turns, with an average of over 130 words per turn. Fig. 3 illustrates the frequencies of selected keywords related to reflective learning. For comparison, additional trials were conducted using only the first paragraph of the prompt to simulate an inexperienced tutor. These dialogues averaged approximately 100 words per turn, indicating a shallower level of discussion. It should be noted that, unlike LLMs, real students may not provide such detailed feedback to a tutor's questions, and more interactive exchanges can be expected with an increased number of dialogue turns. In cases where students were manually simulated to provide mainly simple or partial responses, the dialogues often extended beyond 10 turns.

TABLE III. Key Entity Relationships in Reflective Sessions

| Source | Target | Relationship |
|---|---|---|
| Project-based learning | Team collaboration | Involves |
| Technical skills | Idea implementation | Support |
| Time management | Project timeline | Affects |
| Team role assignment | Team efficiency | Improves |
| Disagreements | Team communication | Pose challenge to |
| Shared vision | Team success | Essential for |
| Conflict resolution | Effective teamwork | Contributes to |
| Reflection | Future project improvement | Leads to |

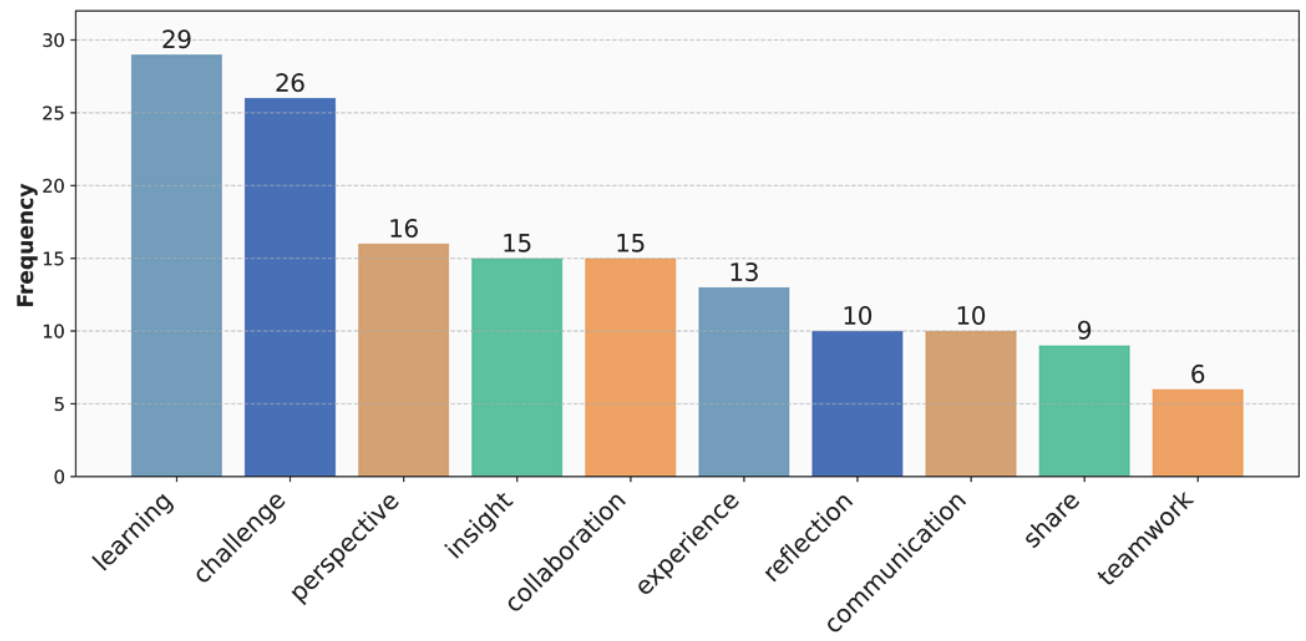

Fig. 3. Distribution of popular keywords across reflective dialogue sessions.

Rubrics are widely employed to evaluate reflective learning outcomes [16]. In this study, GPT-4o was used as an automatic evaluator [17] to analyze a selected dialogue session against two criteria. The findings are summarized in Table IV.

TABLE IV. Overall Quality of Reflective Learning

| Criteria | Description | Score |
|---|---|---|
| Depth of Reflection | The student reflects on challenges thoughtfully, recognizing the importance of collaboration and agreeing on a shared vision. The dynamics and value of disagreements are explored. Further emotional reflection could deepen the analysis. | **8/10** |
| Insights & Learning Outcomes | Clear insights are gained regarding the significance of collaboration, technical skills, and handling conflicting ideas. The example of how differing perspectives improved the project is well explained. Reflection on applying these insights in future contexts could be expanded. | **9/10** |

For *Depth of Reflection*, the session scored 8/10, suggesting that the AI tutor effectively guided the student in exploring the dynamics of experiences. The student reflected on both the challenges that the team overcame and those that remained unresolved, particularly in the context of team collaboration. For *Insights & Learning Outcomes*, the session scored 9/10, indicating that the student gained clear and valuable insights into project-based learning, especially regarding the importance of collaboration and the integration of diverse perspectives.

## B. Student Reflection Analysis

Evaluating the depth of student reflections is essential for understanding how learners engage with and internalize their learning experiences. Such evaluation offers educators valuable insights into the cognitive and metacognitive strategies that students employ when analyzing and applying knowledge. This understanding not only informs the design of targeted interventions to foster higher-order thinking but also guides instructional practices that promote deeper engagement with learning content. Table V presents an analysis of student reflections from a selected session, categorized according to Bloom's taxonomy of learning objectives [18], and illustrates the varying levels of reflective depth across responses.

TABLE V. ANALYSIS OF STUDENT REFLECTIONS BY BLOOM'S TAXONOMY LEVELS

| Response | Bloom's Taxonomy Level | Reflection Depth |
|---|---|---|
| Response 1 | Understanding/ Analyzing | Recognizes team alignment issues and the importance of prioritization; shows comprehension of collaborative dynamics. |
| Response 2 | Analyzing/ Evaluating | Reassesses project-based learning by balancing vision alignment with technical skills; reflects growth in understanding collaboration. |
| Response 3 | Evaluating/ Creating | Challenges assumptions about teamwork, concluding that discussions strengthen ideas; synthesizes new insights into managing disagreements. |
| Response 4 | Creating | Formulates forward-looking strategies (e.g., open-mindedness, early goal alignment), transforming insights into actionable plans and demonstrating a high level of reflective depth. |

- **Response 1**: "One challenge we overcame was the issue of deciding on a final idea because everyone had different visions for what kind of device would be most effective. Initially, it felt like no one would compromise, but eventually, we all managed to align on a concept by prioritizing features that had the most impact on sustainability."
- **Response 2**: "I think I realized that project-based learning isn't just about creating a product; it's also about navigating different perspectives and figuring out how to work together effectively. Before, I used to think that projects were mostly about division of labor, but now I see that even agreeing on the vision is a big part of the process. I also learned that without technical skills, a good idea can only go so far."
- **Response 3**: "Before, I thought it was enough for each person to just handle their part and then bring everything together. I didn't think as much about the importance of agreeing on the big picture early on. This time, because we spent so much time discussing and disagreeing, I realized it wasn't just a 'waste of time,' as I initially thought. It actually helped us see what each person valued, and that made the final idea stronger."
- **Response 4**: "I think I'd try to keep an open mind earlier on in discussions and see if conflicting ideas can enhance the project, rather than viewing them as obstacles. I'd also like to set aside time at the start to make sure everyone is on the same page about the project's goals, rather than assuming we're all aligned from the beginning."

## C. Tutor Guidance Analysis

A sentiment analysis [19] of the tutor across all sessions using GPT-4o revealed that *positive* and *supportive* sentiments were predominant. This reflects the tutor's emphasis on encouragement and empathy to enhance student confidence. *Neutral* sentiments were also frequent, indicating a focus on instruction and summarization. No *critical* sentiments were observed, suggesting a preference for constructive and affirmative feedback. Furthermore, the effectiveness of the tutor's guidance was examined through an analysis of the tutor's prompts from the same session discussed earlier.

TABLE VI. EVALUATION OF TUTOR PROMPTS FOR GUIDING REFLECTION

| Prompt | Effectiveness in Guidance | Facilitation of Learning |
|---|---|---|
| Prompt 1 | Sets a constructive tone by inviting a balanced reflection on both successes and unresolved issues. | Supports the identification of strengths and areas for improvement, fostering a holistic approach to reflection. |
| Prompt 2 | Broadens reflection by linking specific experiences with overarching insights into project-based learning. | Promotes understanding of experiential learning by encouraging the generation of transferable insights. |
| Prompt 3 | Strengthens self-awareness by aiding comparison across experiences and recognition of learning development. | Facilitates deep self-reflection by guiding the student to articulate and appreciate personal growth. |
| Prompt 4 | Encourages the proactive application of insights to future projects, fostering forward-looking reflection. | Cultivates a proactive mindset that encourages the student to transfer insights into new learning contexts. |

The evaluation results in Table VI focus on two primary dimensions. *Effectiveness in Guidance* reflects the tutor's ability to direct students' thinking, encourage balanced reflections, and scaffold the movement toward deeper understanding and application of learning. *Facilitation of Learning* assesses how effectively the tutor supports students in identifying strengths and areas for improvement, thereby fostering self-awareness, personal growth, and the capacity to transfer learning across contexts. Together, these two dimensions capture both the process of effective guidance and the outcomes of facilitated learning, providing a comprehensive perspective on the tutor's overall contribution to reflective practice.

- **Prompt 1**: "Let's start with the big picture. Thinking back on the design of smart devices for campus environmental protection, what's one challenge that you and your team managed to overcome, and one that you feel remained unresolved by the end?"
- **Prompt 2**: "Now, thinking about both of those challenges, what new insights do you feel you've gained about project-based learning through this experience?"

- **Prompt 3**: "I'm curious, though, can you compare this perspective to how you felt about project work before this experience? How did you arrive at these new insights?"
- **Prompt 4**: "Now, looking ahead, how do you think these insights could influence your approach to future projects? Are there specific practices you'd want to carry forward?"

## V. CONCLUSION

This study pioneers the use of generative AI to foster deep and meaningful reflection through interactive dialogue. By adopting a realistic teaching scenario in which a student reflects on a completed team project, it demonstrates the potential of LLMs, specifically GPT-4o, to effectively scaffold reflective processes. The AI tutor's collaborative, reflective, and empathetic guidance enabled the student to generate valuable insights into teamwork, project-based learning, and personal growth through thoughtful questions and constructive feedback. These findings underscore the promise of AI-driven tutoring for supporting reflective practice and provide a foundation for future research into comprehensive frameworks of AI-supported reflective and experiential learning.

Since this study was conducted within a simulated scenario on a specific topic, future research should aim to validate these findings across a broader range of authentic educational contexts. Such large-scale studies are essential to account for the variability among students and instructional models, including differences in students' needs, cultural and linguistic backgrounds, and group dynamics [20], all of which may substantially influence the effectiveness of AI-driven guidance. Moreover, while LLMs can provide objective and consistent evaluation outcomes, future work should prioritize assessment methods that align more closely with learners' lived experiences, perceptions, and situated learning practices, thereby ensuring both validity and educational relevance. Importantly, the role of LLMs as evaluators should be viewed as a methodological complement rather than a replacement for human expertise.

While generative AI demonstrates considerable promise in supporting reflective learning, several critical concerns must be addressed. A key risk is that students may become overly reliant on these systems [21], potentially impeding the cultivation of independent reasoning, critical thinking, and the self-awareness required for effective collaboration. In addition, the risks of bias and hallucinations in AI-generated responses [22, 23], arising from limitations in training data and model architecture, may inadvertently shape student reflections in unintended or unproductive ways. Addressing these challenges is imperative to fully harness the potential of LLMs in reflective learning, while safeguarding the development of independent, critical, and collaborative thinkers.